# Coupling of guided Surface Plasmon Polaritons to proximal self-assembled InGaAs Quantum Dots

G. Bracher[a], K. Schraml, M. Blauth, C. Jakubeit, K. Müller, G. Koblmüller, M. Bichler, M. Kaniber and J.J.Finley

*Walter Schottky Institut and Physik Department, Technische Universität München,*
*Am Coulombwall 4, 85748 Garching, Germany*

**ABSTRACT**

We present investigations of the propagation length of guided surface plasmon polaritons along Au waveguides on GaAs and their coupling to near surface InGaAs self-assembled quantum dots. Our results reveal surface plasmon propagation lengths ranging from $13.4 \pm 1.7$ µm to $27.5 \pm 1.5$ µm as the width of the waveguide increases from 2-5 µm. Experiments performed on active structures containing near surface quantum dots clearly show that the propagating plasmon mode excites the dot, providing a new method to spatially image the surface plasmon mode. We use low temperature confocal microscopy with polarization control in the excitation and detection channel. After excitation, plasmons propagate along the waveguide and are scattered into the far field at the end. By comparing length and width evolution of the waveguide losses we determine the plasmon propagation length to be $27.5 \pm 1.5$ µm at 830 nm (for a width of 5 µm), reducing to $13.4 \pm 1.7$ µm for a width of 2 µm. For active structures containing low density InGaAs quantum dots at a precisely controlled distance 7-120 nm from the Au-GaAs interface, we probed the mutual coupling between the quantum dot and plasmon mode. These investigations reveal a unidirectional energy transfer from the propagating surface plasmon to the quantum dot. The exquisite control of the position and shape afforded by lithography combined with near surface QDs promises efficient on-chip generation and guiding of single plasmons for future applications in nanoscale quantum optics operating below the diffraction limit.

## 1. Introduction

The ability of surface plasmon polaritons (SPPs) to tightly confine electromagnetic fields to nanometer dimensions, combined with advances in nano-fabrication technologies has led to renewed interest in exploring plasmonics for various applications[1] ranging from nano-lithography[2] and microscopy[3], to photovoltaics[4], bio-sensing[5] and plasmonic circuits[6-9]. However, the field enhancements and propagation distances associated with SPPs are limited due to Ohmic losses[10]. The study of the interaction of surface plasmons with active media, a field termed "active plasmonics" by Krasarvin et al. in 2003[11] offers many solutions. For example, it promises to widen the sphere of applicability of surface-plasmon-based applications[12-14] whilst facilitating the study of the emission characteristics of optical emitters in plasmonic systems[15-17]. Most recently, the potential to use proximal active media to provide gain and stimulated emission and amplification of

---

[a] Corresponding author: gregor.bracher@wsi.tum.de; phone +49 89 289 11596

SPPs have now been demonstrated in the optical regime[18-19]. One potentially exciting application could be the possibility to couple discrete quantum emitters with plasmonic nano-devices to realize nanometer scale quantum optical devices[15] that combine quantum light sources with on chip devices such as beam splitters and single photon detectors[20]. Here, we present optical investigations of lithographically defined Au waveguides (WG) on GaAs substrates and probe their coupling to self-assembled near-surface InGaAs QDs. In order to control the distance between the WGs and the QDs, we etch away parts of the 120 nm capping layer with nanometer precision, resulting in 1 mm broad terraces onto which plasmonic structures can be lithographically defined. Metallic waveguides are lithographically defined on these terraces using electron-beam lithography. Although chemically synthesized silver nanowires show remarkably low losses[21-24] and the coupling between these nanowires and colloidal quantum dots has been already been demonstrated[21] the use of lithographically defined structures is expected to provide a higher degree of design flexibility. This is expected to provide limitless possibilities for defining plasmonic nanostructures with enhanced functionality. Examples would include the realization of beam splitters[20] and interferometers[25] to allow one to perform chip based integrated plasmonics. For the optical characterization, we use a low temperature micro photo-luminescence setup with full control of the polarization in both the excitation and detection channels. This set up facilitates the investigation of SPP excitation and propagation, with and without quantum dots at temperatures from 10-300 K.

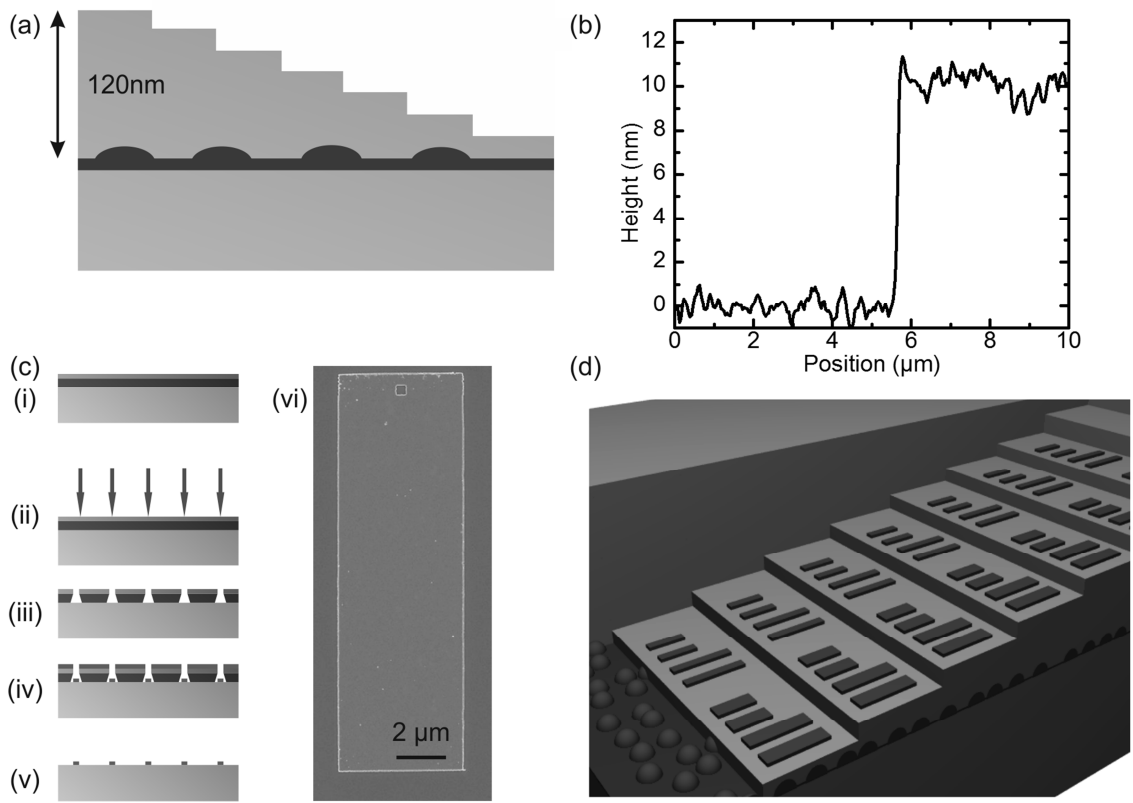

**Figure 1** (a) Schematic representation of the wet chemically etched samples investigated (b) Atomic Force Microscope image of a step: each step has a height of 10 ± 1 nm and a RMS roughness < 1 nm (c) Electron beam lithography process: (i) Spin-coating of bi-layer resist (ii) Exposure with 480 µAs/ cm² (iii) Development (iv) Metallization (v) Lift-off (vi) Scanning Electron Microscope image of a typical 5 µm waveguide (d) Schematic of sample: on each step, waveguides with different lengths and widths are fabricated

## 2. Fabrication

We start our fabrication with an $In_{0.4}Ga_{0.6}As$/GaAs quantum dot sample grown using molecular beam epitaxy. The InGaAs QDs are grown on GaAs substrate using Stranski–Krastanov growth and they are overgrown with a GaAs capping layer. The sample was grown on a nominally undoped GaAs (100) substrate. On top of the substrate a 300 nm thick GaAs buffer was grown at a temperature of 622 °C, followed by a 140 nm thick GaAs layer at 560 °C. At the same temperature 3.5 monolayers of nominally x = 50% $In_xGa_{1-x}As$ were deposited. This resulted in hemispherically shaped self-assembled quantum dots with 25 nm in diameter and 5 nm in height. AFM measurements revealed a density of 120-130 dots per µm². Finally, the sample was overgrown by a 120 nm thick capping layer. In order to control the QD-WG spacing, parts of the capping layer are subsequently removed using citric acid (citric acid : $H_2O_2$ : $H_2O$ = 1 mg : 1 ml : 1 ml ) with an etching rate of 12 nm per minute. Each terrace of the staircase like structure has a width of 1 mm. Figure 1 (b) shows an Atomic Force Microscope (AFM) measurement, from which the step height of 10 ± 1 nm and an RMS roughness of < 1 nm is determined. Figure 1 (c) shows the standard electron beam (e-beam) lithography process for the fabrication on each terrace using standard: a bi-layer e-beam resist is spin-coated using PMMA 200K and PMMA 950K. The structures are written with a JEOL JSM-6400. After development, a 100 ± 5 nm thick gold layer is evaporated by an standard e-beam evaporator and finally the resist is lifted off. Figure 1 (c) (vi) depicts a typical 5µm wide and 25 µm long Au-WG, which is used for measurements. On the upper part one can see a lithographically defined 0.5 µm square aperture that is used to launch SPPs from the far field[26-27]. Figure 1 (d) shows a schematic overview of the completed sample; on each terrace, there are groups of 5 waveguides with the same length and width. The separations of the proximal QDs from the WGs are 7, 17, 27, 37, 47 and 57 nm for the different terraces, respectively.

## 3. Experimental Setup

For our measurements we use a low temperature confocal photoluminescence setup with full control of the polarization in both the excitation and detection channel. Figure 2 shows a schematic of the setup: light from a tunable Ti:Sa laser is polarized using a linear polarizer. A 50:50 beam splitter is used to guide the laser via a 50 x microscope objective with NA = 0.55 to the sample and simultaneously monitor the intensity, which is also used for optical excitation. The light from the sample is collected via the same objective whereupon it passes the beam-splitter, can be analyzed either via a highly sensitive imaging CCD camera or a spectrometer. For the studies of the quantum dot luminescence an additional 900 nm long pass filter is placed in front of the imaging CCD camera to suppress the laser and image the emission from the quantum dots only. A typical integration time for the CCD camera of 5 s is sufficient to image QD luminescence using the excitation power $P_{exc}$= 5 kW/cm².

## 4. Determination of the Propagation Length

For the studies of the propagation length $L_{SPP}$[28] we investigate reference structures lithographically defined on a GaAs wafer with no nearby QDs. To measure $L_{SPP}$ we record the intensity at the end of the WG as a function of the waveguide length. Figure 3 (a) shows on the left hand side an SEM image of a typical 5 µm wide and 45 µm long waveguide. On the right hand side, CCD images of waveguides with different lengths are illustrated: The length increases from (i) to (vii) in steps of 5 µm from 15µm to 45µm. One can see that scattering of SPPs into the far-field is limited to the end of the waveguide. Furthermore, one can see that the intensity reduces with increasing waveguide length. Figure 3 (b) shows a plot of the intensity as a function of the waveguide length at $\lambda_{exc} = 830$ nm. Using an exponential decay, one can extract the propagation length $L_{SPP} = 27.5 \pm 1.5$ µm for 5 µm wide WGs excited at $\lambda_{exc}$= 830 nm. Figure 3 (c) shows $L_{SPP}$ measured for several different WG widths excited at 830 nm, 870 nm and 920 nm. One can see, that $L_{SPP}$ is reduces from

$34 \pm 7$ µm for $w_{wg} = 5$ µm to $15 \pm 1$ µm for $w_{wg} = 2$ µm at $\lambda_{exc}$=870 nm. This observation has been reported in several other experiments[29-30] using scanning optical near-field microscopes. The reduction of the propagation length can be explained by a reduction of the number of supported modes. One can extrapolate the data points, which shows that $L_{SPP}$ will be a few µm or less if the WG width is reduced to sub-micrometer scales. Furthermore, we find that $L_{SPP}$ increases with increasing wavelength. Figure 4 (d) shows a detailed study on the wavelength dependence. We find a good qualitative agreement with the analytical solution of an infinite metal film using the dispersion relation[31] and material constants of gold in the near infrared [32], which is depicted via the dashed line. The values, which we measure, are 40-60% below the expected values. The discrepancy can be explained by surface roughness and grain boundaries resulting from the evaporation process and by the finite WG width. Nevertheless, all waveguide widths are suitable for broadband waveguiding in the near infrared. The waveguides cover the bandgap of GaAs at low temperatures as well as the spectral emission from the quantum dots. Therefore, they are suitable to guide light, which is used for the excitation of the QDs as well as the emission of the QDs.

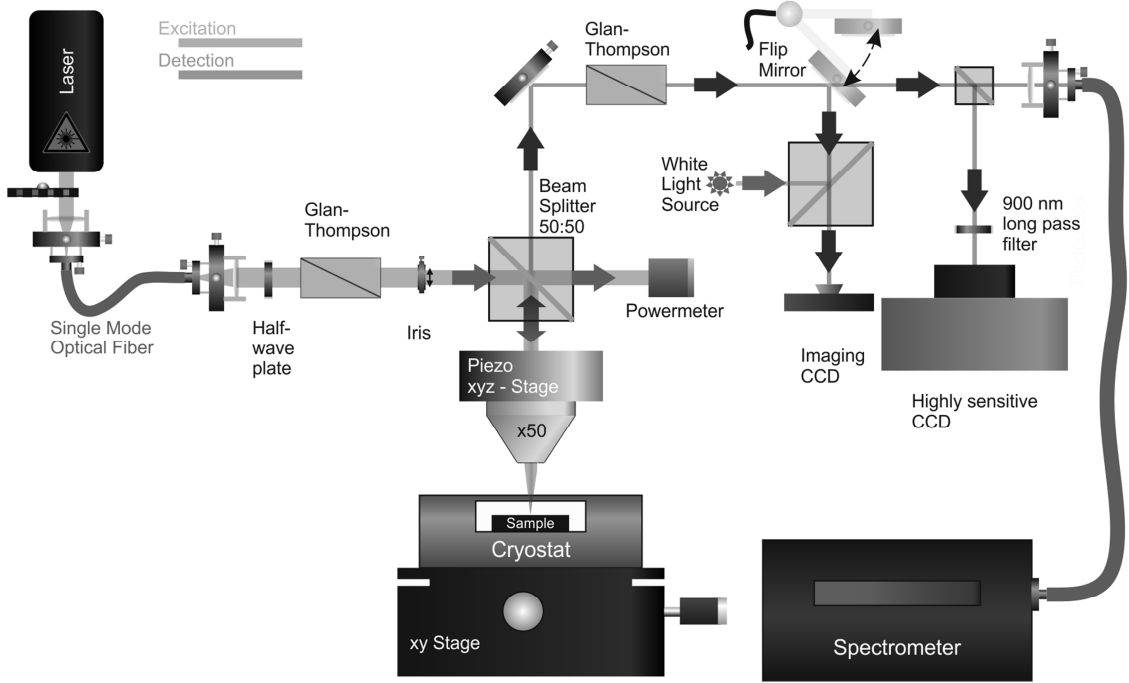

**Figure 2** Setup: light from a tunable Ti:Sa laser is polarized and focused via a beam-splitter and a 50 x microscope objective with NA=0.55 onto the sample. The light from the sample is collected via the same objective. A second linear polarizer is used to suppress the laser. The rough positioning is done with a normal CCD. For the measurements, either a highly sensitive CCD with an optional 900 nm long-pass filter or a spectrometer can be used.

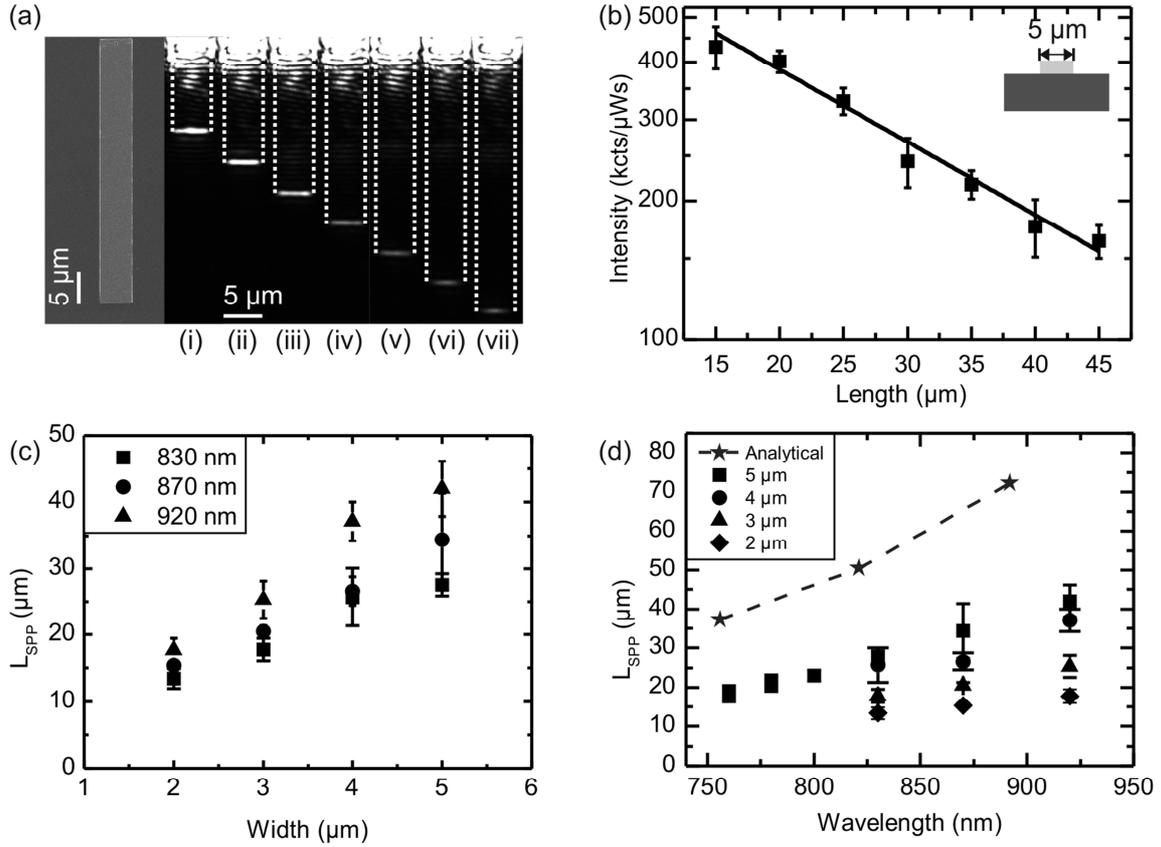

**Figure 3** (a) left: SEM image; right: Images obtain by the CCD camera- length of WGs increase in pictures (i)-(vii) from 15-45 µm in steps of 5 µm (b) Intensity plotted as a function of the WG length (c) Propagation length as a function of the waveguide width plotted for different excitation wavelengths (d) Wavelength dependence of propagation length plotted for various waveguide widths

## 5. Excitation of Quantum Dots

In order to study the photo luminescence emission from the near surface QDs we use a 900 long pass filter in order to suppress the excitation laser at $\lambda_{exc}$=819 nm. The wavelength of the excitation laser was chosen to generate electron-hole pairs in the bulk GaAs close to the band edge. Therefore, the diffusion length of the charge carriers is limited and they are rapidly trapped over picosecond timescales by QDs in the vicinity of the waveguide[33]. Figure 4 (a) shows a typical CCD image recorded from the WG end. One can see that the waveguide is surrounded by QD luminescence. In addition, we observe very strong luminescence at the end of the waveguide compared to the WG edges. This is in good agreement with the emission pattern from the waveguide studies, where we find that the far-field emission is limited to the end of the waveguide. Therefore, we expect also the strongest excitation of QDs where the out-coupling is most efficient. For the detailed analysis of the QD luminescence we use a spectrometer. Figure 4 (b) shows two typical spectra measured at the end of the WG, where one can see luminescence from several dots. The black curve shows the spectrum for an excitation polarization along the WG and the gray curve for a polarization perpendicular to the WG axis. One can see, that the intensity is reduced by 50%, when excited perpendicular to the WG axis. Figure 4 (c) shows a quantitative investigation of this behavior. We integrated the spectrum from 880 nm to 950 nm and plotted the intensity as function of the orientation of the polarization. We find a degree of polarization of DoP= $(I_{max}- I_{min})/( I_{max}+ I_{min}) = 50 \pm 2$ %. Figure 4 (d) depicts the polarization dependence in the detection channel. Here, we find a degree of polarization with a DoP = 12 ± 2%. Comparing these two values to a previous work[28], we conclude that the excitation of the QD is indirect, SPPs generating electron hole-pairs in the semiconductor whilst propagating along the waveguide.

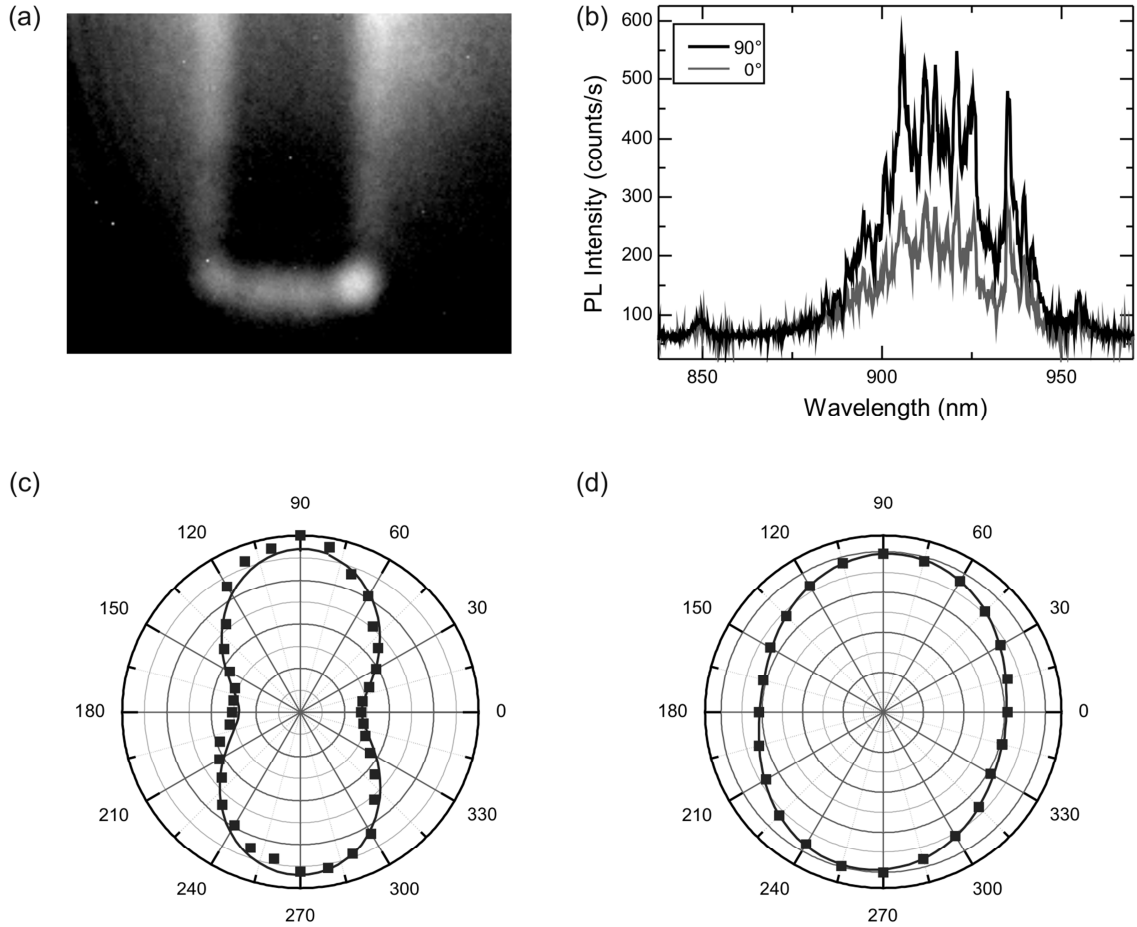

**Figure 4** (a) CCD image using 900 long pass filter: WG are surrounded by QD luminescence (b) PL spectrum for different excitation polarizations (c) Polarization dependence of excitation: SPPs are generated most efficiently along the waveguide axis (d) Polarization dependence of detection: QDs have no preferred emission polarization

## 6. Conclusion

In summary, we have reported the fabrication and optical investigation of hybrid Au-GaAs plasmonic nanostructures fabricated using electron beam lithography. Both passive and active structures were investigated that did or did not contain proximal InGaAs self-assembled quantum dots. For rectangular SPP waveguides defined on GaAs we measured the surface propagation length $L_{SPP}$ as a function of the waveguide width and the excitation wavelength. Good qualitative agreement is found with the theoretically expected behavior, a reduction of $L_{SPP}$ from 34 ± 7 μm to 15 ± 1 μm having been observed upon narrowing the waveguides from 5-2 μm. This reduction was attributed to a reduction of the number of supported SPP modes upon narrowing the waveguides. Finally, in active structures containing proximal quantum dots we have demonstrated the excitation of near-surface self-assembled QDs using SPPs. The QD luminescence is found to be unpolarized with a DoP = 12 ± 2%, whilst the excitation shows a strong preference along the WG axis with 50 ± 2 %. The excitation of near surface self-assembled QDs via SPPs opens the door for future on-chip quantum optics experiments.

## Acknowledgments

We acknowledge financial support of the DFG via the SFB 631, Teilprojekt B3, the German Excellence Initiative via NIM, and FP-7 of the European Union via SOLID.